# “Self-controlled growth”, “Coherent shrinkage”, “Eternal life in a self-bounded space” and other amazing evolutionary dynamics of stochastic pattern formation and growth models inspired by Conway’s “Game of Life”

Leonid Yaroslavsky,

Dept. of Physical Electronics, School of Electrical Engineering,

Tel Aviv University, Tel Aviv 69978, Israel. (yaro@eng.tau.ac.il)

## ABSTRACT

Results of experimental investigation are presented of evolutionary dynamics of several stochastic pattern formation and growth models designed by modifications of the famous mathematical “Game of Life”. The modifications are two-fold: “Game of Life’s” rules are made stochastic and mutual influence of neighboring cells is made non-uniform. The results reveal a number of new phenomena in the evolutionary dynamics of the models:

- “Ordering of chaos to maze-like patterns”: evolutionary formation, from arbitrary seed patterns, of stable maze-like patterns with chaotic “dislocations” that resemble natural patterns frequently found in the nature, such as skin patterns of some animals. The remarkable property of these patterns is their capability of unlimited growth, self-healing and transplantation.
- “Self-controlled growth” of chaotic “live” formations into “communities” bounded, depending on the model, by a square, hexagon or octagon, until they reach a certain critical size, after which the growth stops.
- “Coherent shrinkage” of “mature”, after reaching a certain size, “communities” into one of stable or oscillating patterns preserving in this process isomorphism of their bounding shapes until the very end.
- “Eternal life in a self-bounded space” of “communities”: seemingly permanent “birth/death” activity of “communities” after they reach a certain size and shape.

## 1. INTRODUCTION

In this paper we consider evolutional dynamics of several pattern formation and growth models derived through modifications of the famous mathematical model known as Conway's "Game of Life" ([1]). In this model, 2D arrays of binary, i.e. assuming values 1 ("live") or 0 ("empty"), cells arranged in nodes of a rectangular lattice within a rectangular "vital space" of a finite size (see Figure 1) are subjected to evolution.

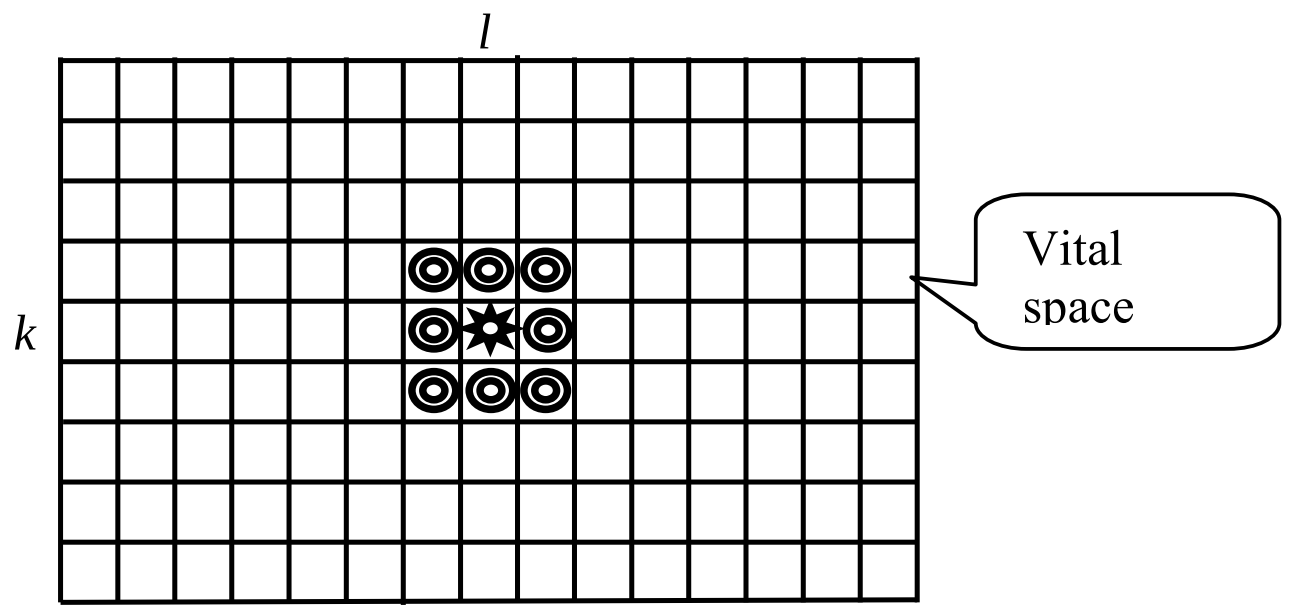

Figure 1. Cells in a "vital space" on a rectangular lattice. Donuts indicate 8-neigboring cells of the cell indicated by the 8-point star

Each cell in this array has 8 neighbors (Figure 1). Rules of the evolution are as following. At each subsequent step of evolution,

(i) each "empty" cell that has exactly 3 "live" neighbor cells in its 3x3 neighborhood gives a "birth", i.e. becomes "live";

(ii) each "live" cell that has less than 2 and more than 3 "live" cells in the neighborhood "dies", i.e. becomes "empty";

(iii) otherwise nothing happens.

Patterns generated by this model in course of evolutionary steps (evsteps) indexed by $\boldsymbol{t}, \boldsymbol{t} = \mathbf{1,2,...}$ can formally be described by the equation:

$$\boldsymbol{pattern}^{(t)}(k,l) = \left\{\boldsymbol{pattern}^{(t-1)}(k,l)\delta\left[S_8^{(t-1)}(k,l)-2\right]\right\} \forall \left\{\delta\left[S_8^{(t-1)}(k,l)-3\right]\right\};$$

$$\boldsymbol{pattern}^{(0)}(k,l) = \boldsymbol{seed_pattern}(k,l) \qquad (1)$$

where $(k,l)$ are integer indices of cells on a rectangular lattice, $\delta(\cdot)$ is the Kronecker delta ($\delta(x) = \mathbf{0}^x$), $S_8^{(t-1)}(k,l)$ is the sum of binary values of 8 neighboring cells of (***k,l***)-th cell on the lattice, $\boldsymbol{seed_pattern}(k,l)$ is an arbitrary initial binary pattern used as a seed, and $\{.\}\forall\{.\}$ denotes element-wise logical "OR" operation on arrays of binary numbers.

The standard "Game of life" model produces, in course of evolution from an arbitrary seed pattern, three types of patterns: (i) stable patterns, which, once appeared, remain unchanged unless they collide with neighbor patterns, which can might happen in course of evolution; (ii) periodical patterns ("oscillators"), which repeat themselves after a certain number of the evolution steps; obviously, above-mentioned stable patterns can be regarded as a special case of periodical patterns with a period of one step; (iii) self-replicating "moving" patterns, or "gilders", which move across the lattice and replicate themselves in a shifted position after a certain number of steps; this can be regarded as a general "space-time" periodicity.

Since its invention, "Game of life" has been intensively studied experimentally by numerous enthusiasts, which have been competing between each other in discovery of new stable patterns and oscillators. As a result, very many types of stable patterns, "oscillators" and "gliders" have been discovered, including very sophisticated ones such as "Gosper Glider Gun" and "2C5 Space Ship Gun P690". These patterns as well as their classification issues and rates of appearance one can find elsewhere ([2]).

"Game of life" is not only a splendid plaything for mathematicians and amateurs. It can also, after appropriate modifications, serve as a base for evolutionary 2D pattern formation and growth models, and, more generally, for models of 2D nonlinear dynamic systems with feedback ([3-6]).

In the present paper, we introduce several new modifications of the standard Conway's model and describe results of computer simulation experiments, which reveal a number of new phenomena in the evolutionary dynamics of the models.

## 2. MODIFICATIONS OF THE STANDARD CONWAY'S "GAME OF LIFE" MODEL

The modifications of the standard Conway's model given by Eq. 1 are made in two ways:

(i) "Deaths" of "live" cells are made stochastic with a certain probability $P_{death} \leq 1$, which is a model parameter.

(ii) Counting the number of "live" cells in 8-neighborhood of each cell by means of summation $S_8^{(t-1)}(k,l)$ of their binary values is replaced by a weighted summation with rounding up the summation result:

$$\tilde{S}_w^{(t-1)}(k,l) = Round\left(\sum_{m=-1}^{1}\sum_{n=-1}^{1} w_{m,n}\, pattern^{(t-1)}(k-m, l-n)\right), \qquad (2)$$

where weights $\{w_{m,n}\}$ are entries of a 3x3 weight matrix $\mathbf{W} = \{w_{m,n}\}$ that specifies the model. Weight matrices selected for our study are presented in

Table 1.

**Table 1**

| | | | | |
|---|---|---|---|---|
| Standard_mask $\begin{bmatrix}1&1&1\\1&0&1\\1&1&1\end{bmatrix}$ | Isotropic_mask $\begin{bmatrix}0.7&1&0.7\\1&0&1\\0.7&1&0.7\end{bmatrix}$ | Diagonal_mask $\begin{bmatrix}1&0.7&1\\0.7&0&0.7\\1&0.7&1\end{bmatrix}$ | Cross_mask $\begin{bmatrix}0.3&1&0.3\\1&0&1\\0.3&1&0.3\end{bmatrix}$ | Cross4_mask $\begin{bmatrix}0&1&0\\1&0&1\\0&1&0\end{bmatrix}$ |
| Cross4diag_mask $\begin{bmatrix}1&0&1\\0&0&0\\1&0&1\end{bmatrix}$ | Hex0_mask $\begin{bmatrix}1&0&1\\1&0&1\\1&0&1\end{bmatrix}$ | Hex1_mask $\begin{bmatrix}0.75&0.5&0.75\\1&0&1\\0.75&0.5&0.75\end{bmatrix}$ | Hex2_mask $\begin{bmatrix}1&0.75&0.5\\0.75&0&0.75\\0.5&0.75&1\end{bmatrix}$ | |

## 3. ORDERING OF CHAOS: THE STANDARD MASK MODEL

We begin with the more or less known type of the evolutionary dynamics, the “ordering of chaos”. By the “ordering of chaos” we mean formation, out of, generally, chaotic seed patterns, stable formations that are “fixed points” of the model, i.e. formations, which, once appeared, are either not changing or oscillating in space or in “time” in course of the evolution.

Obviously, fixed points of the model must be patterns that consist of “live” cells with only two or three “live” neighbors, which would not “die” on the next step of evolution, and of “empty” cells with less or more than three “live” neighbor cells, which would not come to life. Stable and oscillating patterns such as those shown in Figure 2, left image generated by the standard non–stochastic model, i.e. for the probability of “death" $\boldsymbol{P_{death}}$=1, are well known and well reported [2]. However, it turns out that such patterns are rather an exception than the rule because they emerge only when $\boldsymbol{P_{death}}$ is strictly equal to one. When $\boldsymbol{P_{death}}$ is even only slightly less than one, oscillating formations characteristic of the non-stochastic standard Conway’s model occasionally collapse, producing chaotic clouds of “live” and “dying” cells that collide with each other and do not seem stabilizing ever. As $\boldsymbol{P_{death}}$ becomes lower, these clouds are becoming denser in the “vital space” and gradually fill it up keeping their “births/deaths” activity seemingly permanently and demonstrating sort of “eternal life in the vital space”.

Furthermore, our experiments reveal that for the probability of “death” lower than about 0.7, in different parts of the “vital space” patches of maze-like patterns of different size and orientation emerge and grow in the “sea” of “active” chaos (Figure 2, middle image). The “birth/death” activity is becoming concentrated mainly on the borders of these patches and remains to be seemingly permanent while $\boldsymbol{P_{death}}$ is higher than about 0.3. When $\boldsymbol{P_{death}}$ becomes lower than about 0.3, patches’ borders tend, after a certain number of the evolutionary steps, to stabilize and stable "mature" maze-like patterns emerge consisting of patches of alternative vertical and horizontal stripes of "live" and "empty" cells chaotically interrupted by dislocations, in which the direction of stripes is either switched to the perpendicular one or stripes of "live" and "empty" cells switch their positions (Figure 2, right image). These patterns are, obviously, fixed points of the proper Conway's model as well.

| $\boldsymbol{P_{death}} \equiv 1$<br>Stable patterns and "oscillators" | $\approx 0.3 < \boldsymbol{P_{death}} < 1$<br>"Eternal life in the vital space" | $0 < \boldsymbol{P_{death}} < \approx 0.3$<br>Maze-like formations |
|---|---|---|

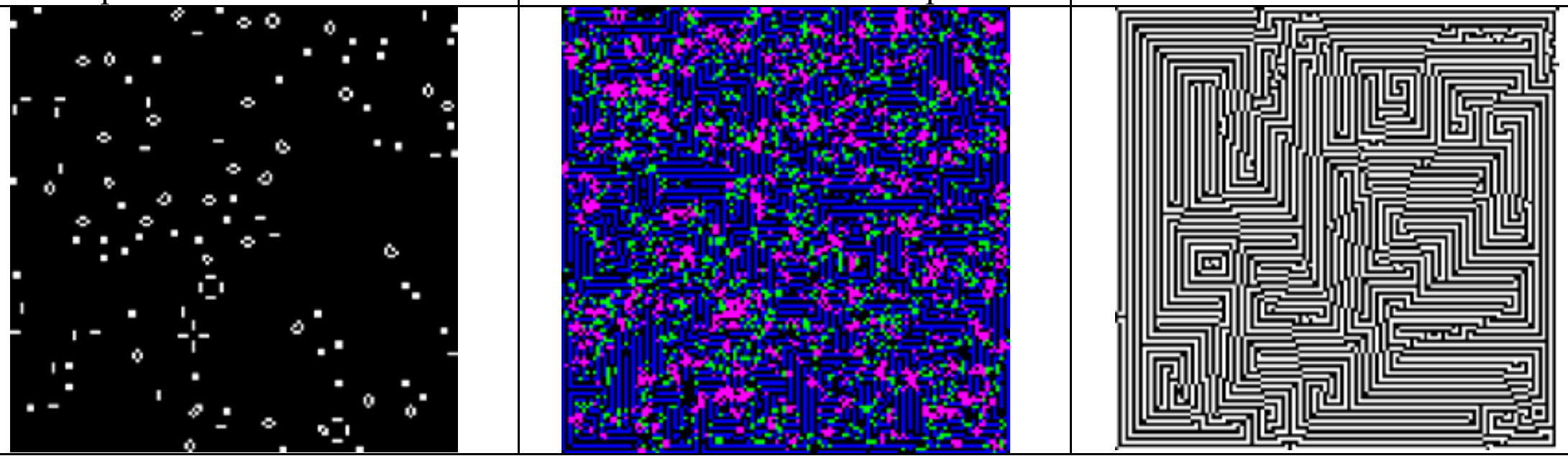

Figure 2. Three modes of the evolutionary dynamics of the stochastic modification of the "standard" Conway's mask model: emerging of individual stable and oscillating formations (for $\boldsymbol{P_{death}} \equiv 1$); emerging stable maze-like pattern (for $0 < \boldsymbol{P_{death}} < \approx 0.3$) and "eternal life in the vital space" (intermediate values of $\boldsymbol{P_{death}}$). In the color coded image, cells that are to "die" are shown pink, cells are to give "birth" are shown green, "live" stable cells are shown blue and empty cells are shown black. In black and white images, "live" cells are shown white.

Emergence of maze-like patterns as fixed points of the standard Conways' model for sufficiently low probability of "death" was already reported earlier ([3-6]). The present experiments reveal a new amazing property of these stable maze-like stable patterns, their potential ability to growing, to self-healing and to transplantation. If one takes, as a seed pattern, a fragment of a maze-like stable pattern or a maze-like stable pattern with a hole and let them evolve with the probability of "death" $\boldsymbol{P_{death}} < \approx 0.3$**,** the former will grow until it fills up the entire vital space and the latter will grow to fill up the hole, as it is illustrated in Figure 3. One can also transplant fragments of one maze-like stable pattern into another and use the pattern with the implanted fragment as a seed pattern for further evolution of the model. After some number of evolutionary steps, transplanted fragment perfectly tailors itself in the new "home" as it is shown in Figure 4.

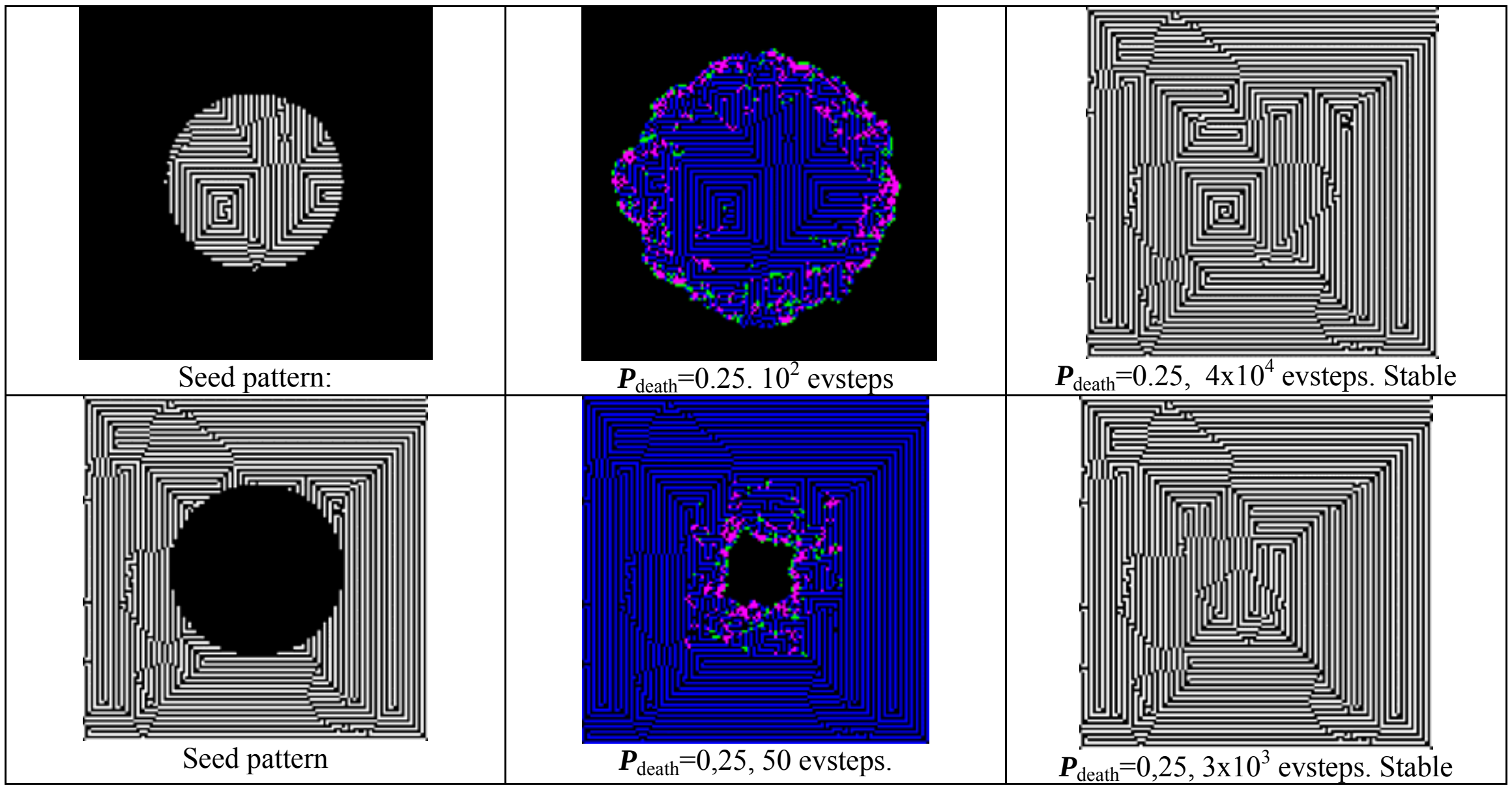

Figure 3. Growth and "self-healing" capability of the maze-like patterns generated by the Standard_mask model. In color coded images, cells that are to "die" are shown pink, cells are to give "birth" are shown green, "live" stable cells are shown blue and empty cells are shown black. In black and white images, "live" cells are shown white.

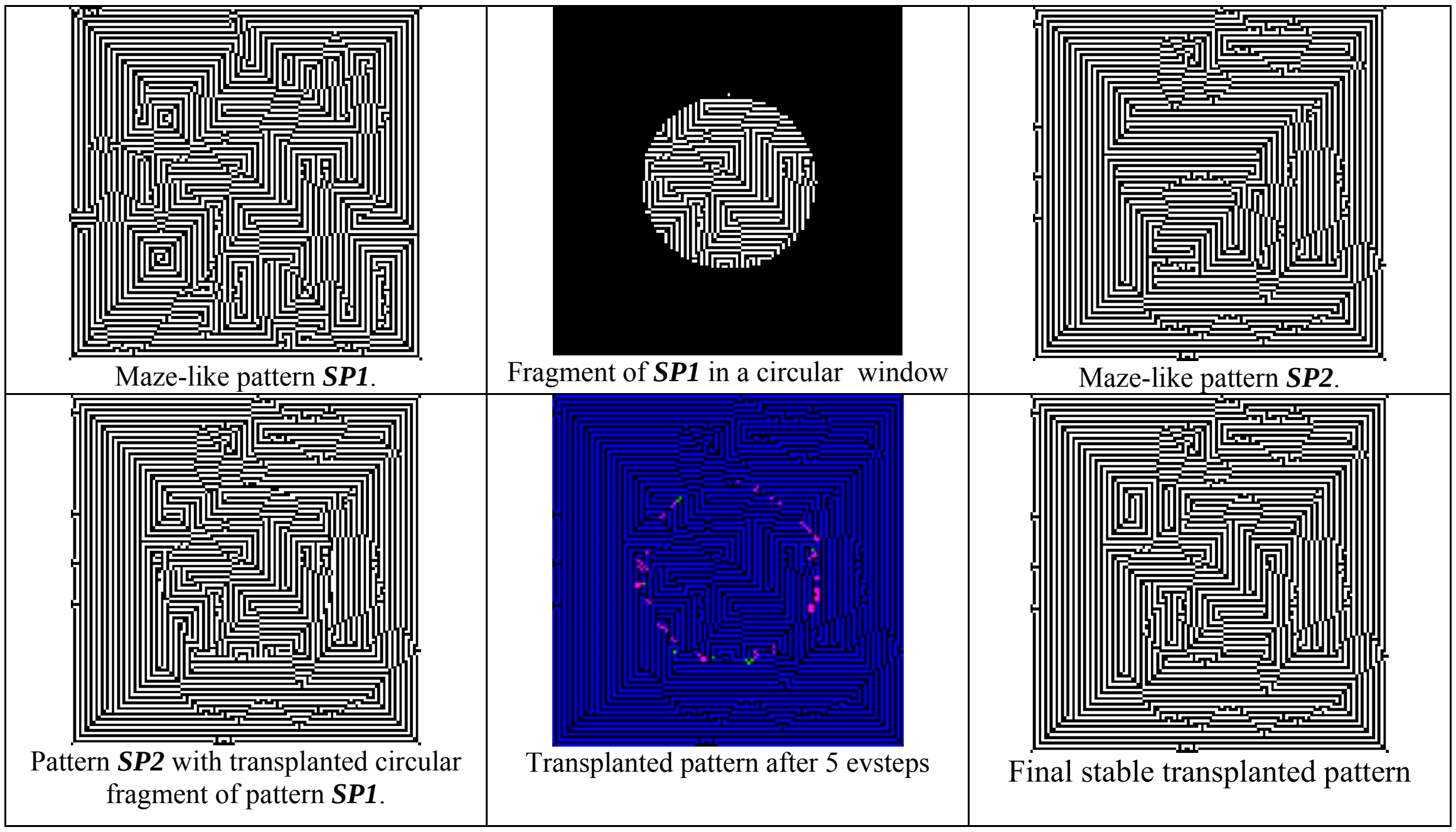

Figure 4. Transplanting a circular fragment of maze-like pattern ***SP1*** into pattern ***SP2*** (both are generated by the Standard_mask model). In color coded images, cells that are to "die" are shown pink, cells are to give "birth" are shown green, "live" stable cells are shown blue and empty cells are shown black. In black and white images, "live" cells are shown white.

## 4. "ORDERING OF CHAOS THROUGH "SELF-CONTROLLED" GROWTH

For $\boldsymbol{P_{death}} \equiv \mathbf{1}$, the other models introduced in Sect. 2 demonstrate "ordering of chaos" dynamics similar to that of the standard stochastic Conway's model: their evolution ends up with a set of isolated stable formations shown, in order to not divert readers' attention, elsewhere ([7]). For $\boldsymbol{P_{death}} < \mathbf{1}$, "ordering of chaos" type of the evolutionary dynamics similar to that for the standard model was observed only for models with "Isotropic"_mask and Hex2_mask: for $\approx \mathbf{0.85} < \boldsymbol{P_{death}} <= \mathbf{1}$, they end up with isolated individual stable formations and for $\mathbf{0} < \boldsymbol{P_{death}} <\approx \mathbf{0.5}$, they converge to stable maze-like patterns (see Figure 5**).**

| **Isotropic_mask** | | **Hex2_mask** | |
|---|---|---|---|

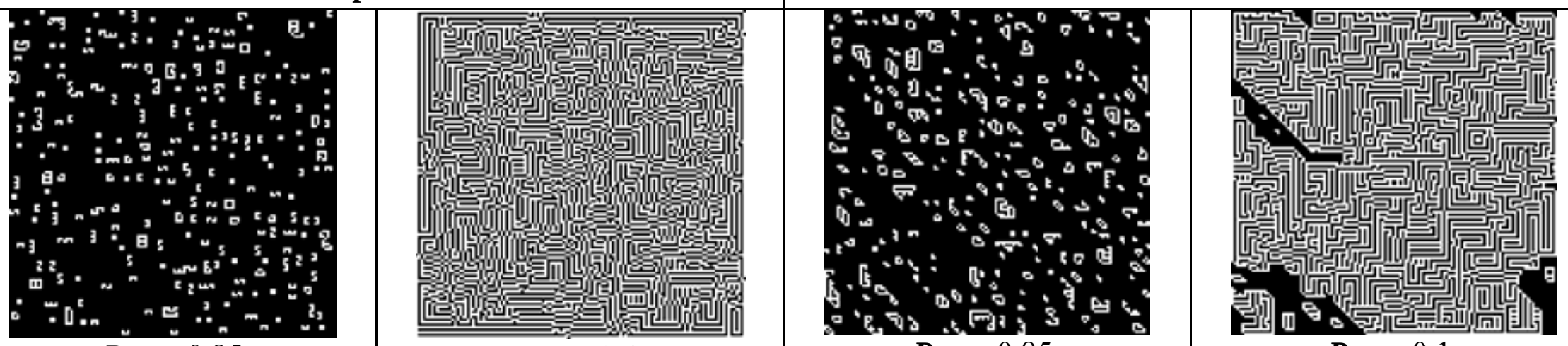

| | | | |
|---|---|---|---|
| $\boldsymbol{P}_{\text{death}}$=0.85 | $\boldsymbol{P}_{\text{death}}$=0.4; | $\boldsymbol{P}_{\text{death}}$=0.85; | $\boldsymbol{P}_{\text{death}}$=0.1; |

Figure 5. Two types of "ordering of chaos" dynamics for Isotropic_mask and Hex2_mask models for $\boldsymbol{P}_{\textbf{death}}$<1. "Live" cells are shown white, empty cells are shown black.

However, in distinction from the standard model, growth of maze-like patterns generated by these modified models has a very special character, it appears as "self-controlled". As one can see from Figure 6, circular fragments of the stable maze-like patterns chosen as seed patterns grow only until growing patterns reach a square, for the Isotropic_mask model, or a hexagon, for the Hex2-mask model, shapes that circumscribe, in case of solid seed patterns, the shape of the seed pattern. Then the growth stops. The self-controlled growth capability of some of the models is reflected also in their "self-healing" capability to fill holes in seed patterns. While isotropic_mask model does fill the hole, as can be seen from Figure 6, Hex2-model fills up the hole only partially leaving empty configuration with horizontal, vertical and diagonal-oriented borders similar to those of the above mentioned hexagon bounded maze-like formation.

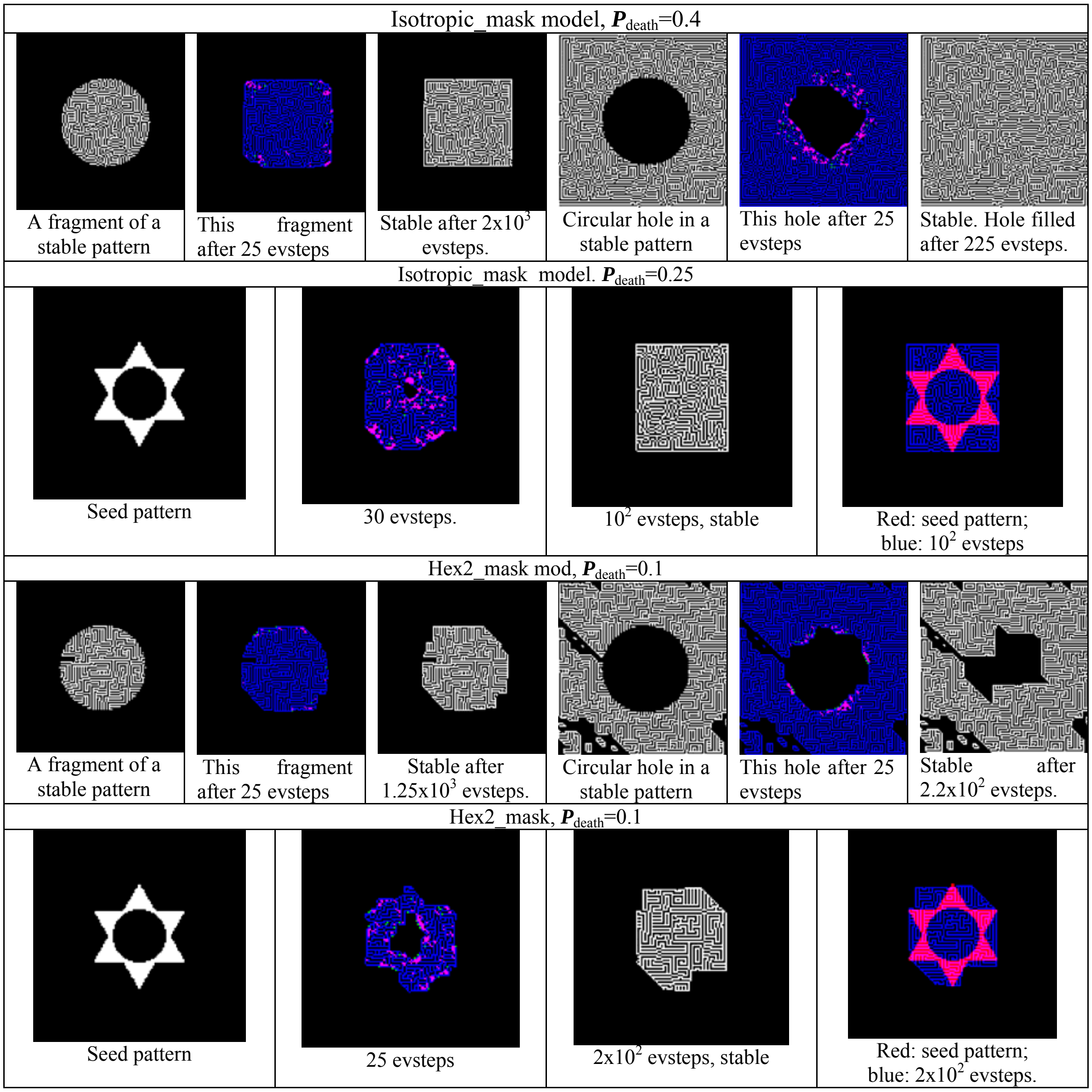

Figure 6. Examples of “self-controlled growth” of maze-like patterns generated by the Isotropic and Hex2_mask model. In color coded images cells that will “die” on the next step are shown pink, cells that will give “birth” are shown green, stable cells are shown blue and empty cells are shown black; in black and white images “live” cells are shown white.

## 5. “COHERENT SHRINKAGE” AND “ETERNAL LIFE IN A SELF-BOUNDED SPACE” MODES OF EVOLUTIONARY DYNAMICS

Perhaps, the most amazing, along with above-mentioned “self-controlled growth”, types of evolutionary dynamics observed in the experiments are “coherent shrinkage” and “eternal life in a self-bounded space”. They were observed with some models for probabilities of “death” in the middle of the range 0÷1. Figure 8 and Figure 9 illustrate the phenomena of “self-controlled growth” and of “Coherent shrinkage” observed for dense chaotic and “solid” seed patterns.

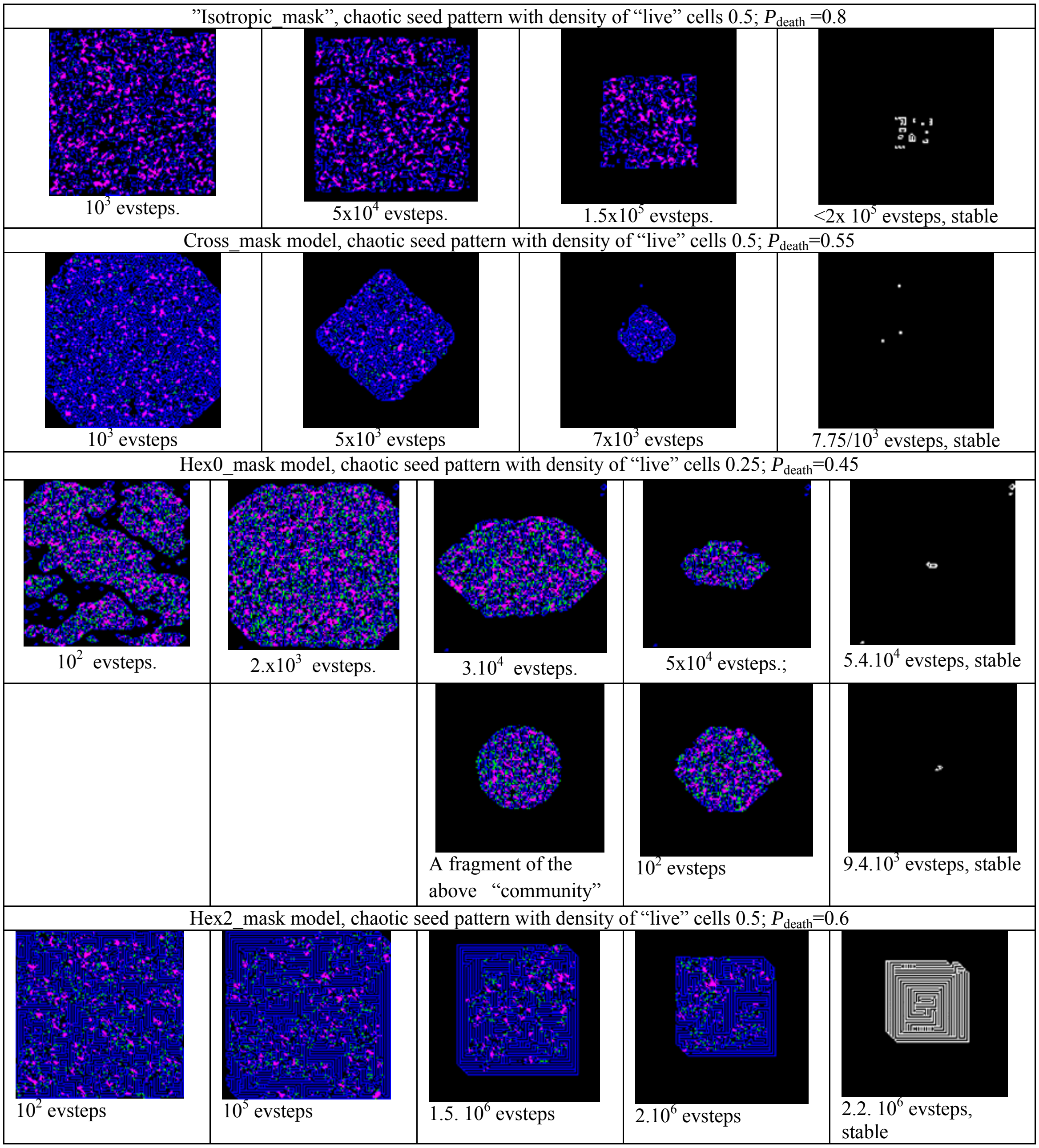

Figure 7. “Self-controlled growth” and “coherent shrinkage” of patterns emerged from chaotic seed patterns for Isotropic_mask, Cross_mask, Hex0- and Hex2_mask models. Images in the fourth from the top row show evolution of a fragment of the hexagon-shaped pattern obtained after $3.10^4$ evsteps planted into an empty space (in color coded images cells that will “die” on the next step are shown pink, cells that will give “birth” are shown green, stable cells are shown blue and empty cells are shown black; in last images of every row, “live” cells are shown white).

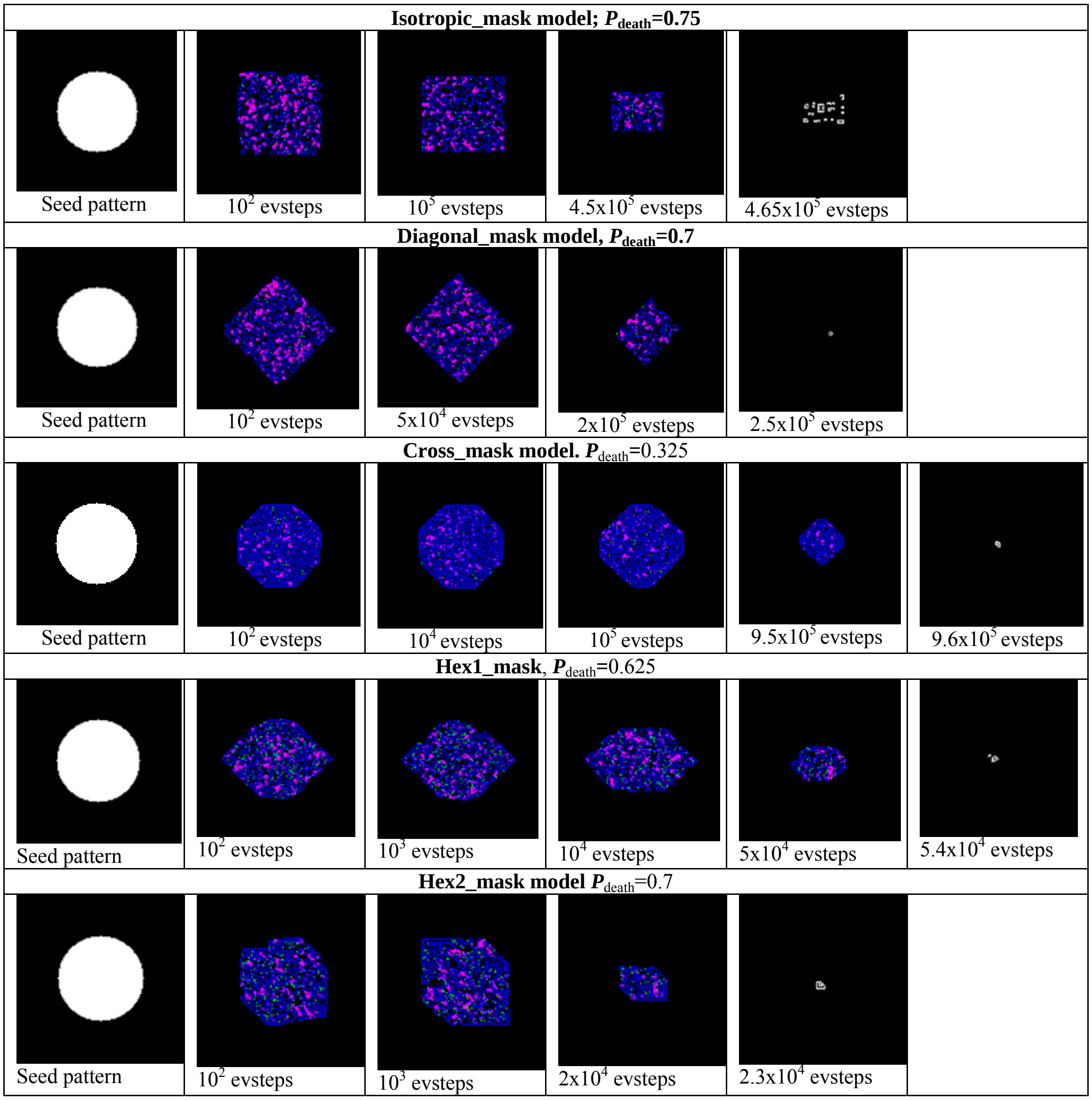

Figure 8. “Self-controlled growth” and “Coherent shrinkage” of patterns emerged from solid seed pattern “Circle”. In color coded images cells that will “die” on the next step are shown pink, cells that will give “birth” are shown green, stable cells are shown blue and empty cells are shown black; in black and white images “live” cells are shown white.

As one can see from the figures, in the “coherent shrinkage” mode the models apparently pass, in course of evolution, through stages of a sort of a “life cycle”:

- “Birth”: loci of growth emerge in seed patterns.
- “Childhood and adolescence”: born formations grow in size, forming “communities” of cells. For chaotic seed patterns, in which “live” cells fill more all less uniformly the entire

"vital space", this growth goes on within the "vital space". For "solid" seed patterns, the growth is "self-controlled": it goes on till "communities" reach a shape bounded, depending on the model, by a square (Isotropic and Diagonal_mask models), an octagon (Cross_mask model) or a hexagon (Hex0, Hex1 and Hex2-mask models). Upon reaching a certain critical size, the emerged shaped communities stop growing further unless they touch another neighbor "community". In this case touching communities merge to form larger "communities", which continue growing till they reach a similarly bounded shape of a larger size. In such a way "communities" reach a state of "maturity".

- The state of "maturity": bounded shaped mature "communities" stay like islands in the "ocean" of empty cells and keep their activity ("births" and "deaths") and their overall size and shape during a certain number of evolution steps, which depends on the probability of death: the lower the probability of death the larger this "population" stability period.
- "Senescence". After a certain period of relative stability in size and shape, the "communities" begin to gradually shrink. The shrinkage appears to be "coherent": the "communities" are coherently shrinking from their borders preserving isomorphism of their shapes till the very end, when they either completely disintegrate to nil or, most frequently, end up with one of stable formations. In this process, "communities" preserve, in the course of the "coherent shrinkage", potentials of growth: as one can see in the third and fourth rows of Figure 7, if one extracts a fragment of a shrinking "community" and plants it into an empty space, the planted fragment resumes growing until it reaches a maturity state in a self-bounded shape, characteristic for the given model; after that it starts shrinking in the same way as its "mother community" does.

Experiments show that "coherent shrinkage" slows down with lowering the probability of "death". For the Diagonal_mask, Isotropic_mask, "Cross_mask", Hex0_mask, and Hex1_mask models, this slowing down might be so substantial that the dynamics of the models appears as kind of “eternal life in a self-bounded space”: upon reaching the state of “maturity”, communities, if they are sufficiently large, stay active (in terms of “births” and “deaths”) and keep their outer bounds during millions of evolution steps and apparently forever. We illustrate this in Figure 9, Figure 10 and Figure 11.

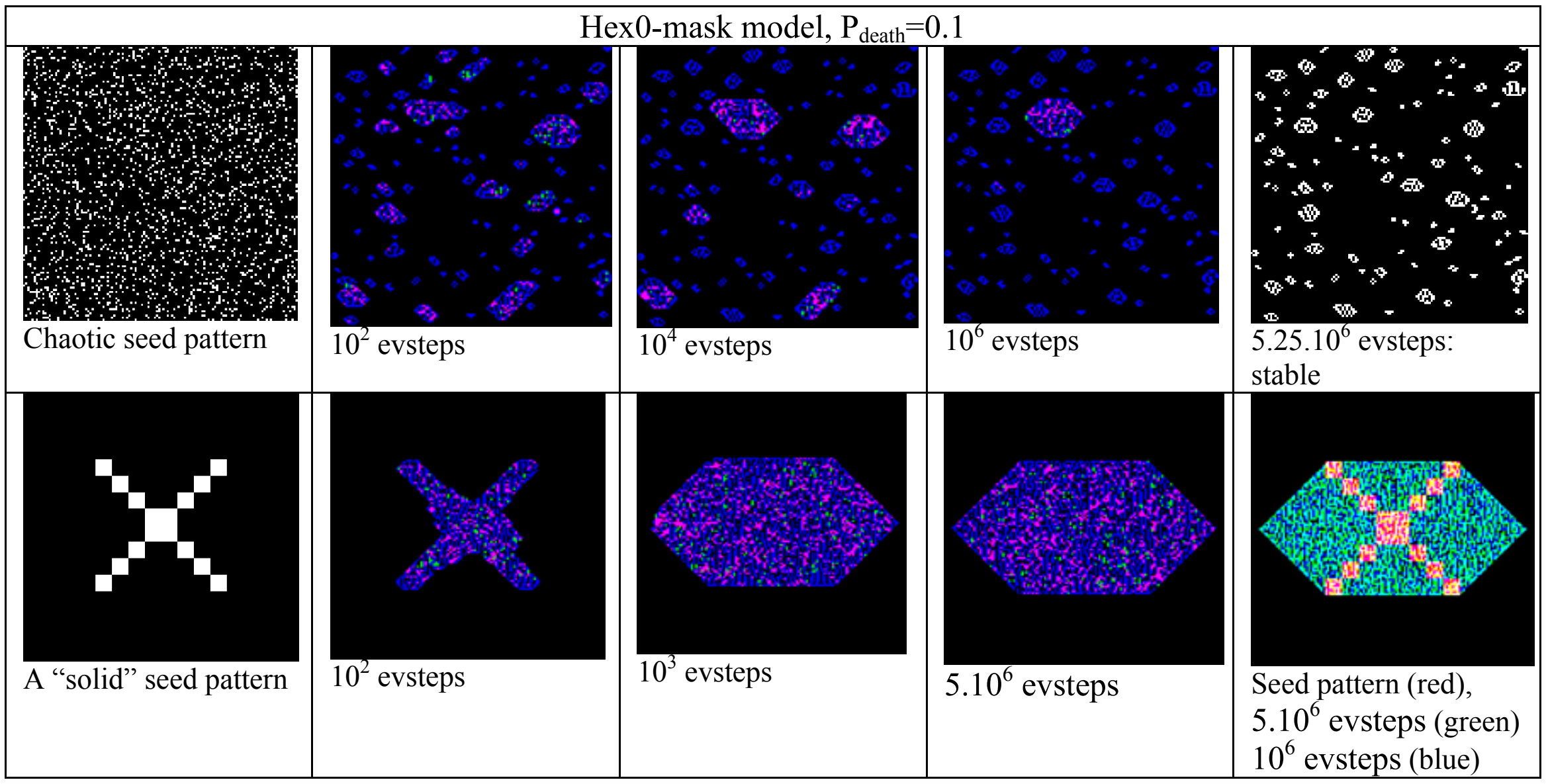

Figure 9. Hex0_mask models: “Coherent shrinkage” and “Eternal life in a self-bounded space” for the same probability of “death”. In color coded images, cells that are to “die” on the next step are shown pink, cells that are to give “birth” are shown green and stable “live” cells are shown blue; “empty” cells are shown black.

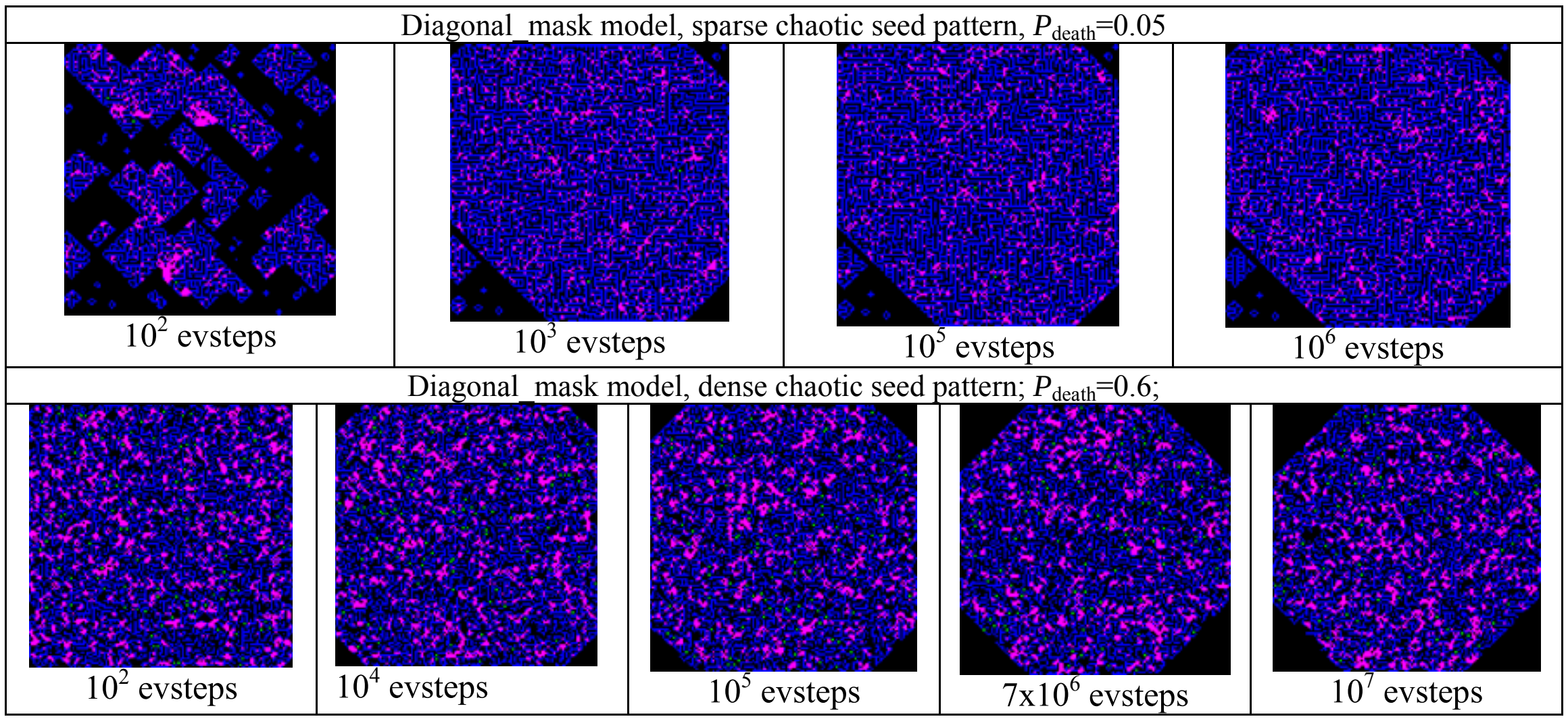

Figure 10. Diagonal_mask model: “eternal life in a self- bounded space for dense and sparse “chaotic” seed patterns with densities of “live” cells 10% and 50% . In color coded images, cells that are to “die” on the next step are shown pink, cells that are togive “birth” are shown green and stable “live” cells are shown blue; “empty” cells are shown black. Pay attention to emergence of empty corners in the vital space.

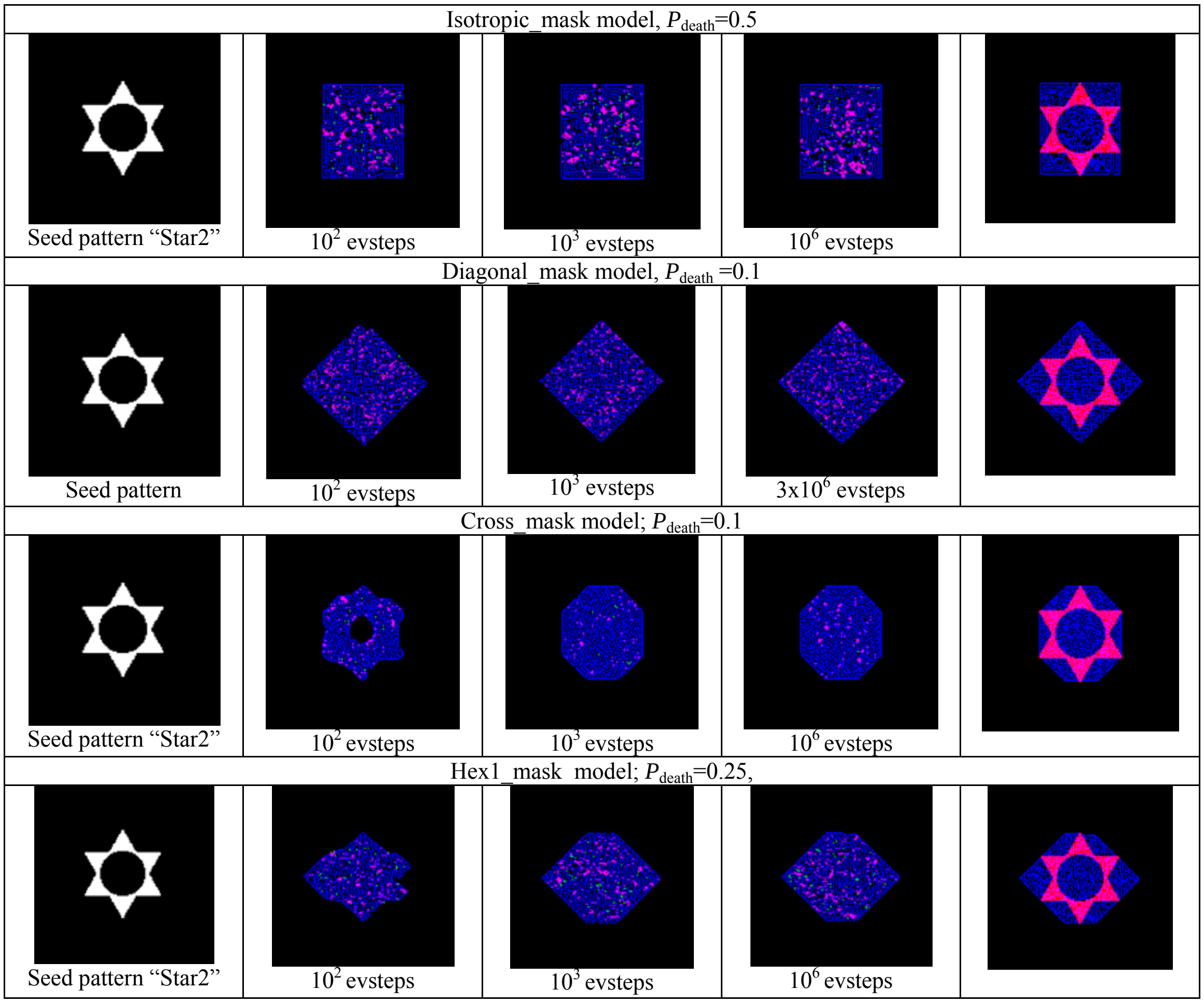

Figure 11. Examples of "eternal life in a self-bounded space" dynamics for one of the used solid seed patterns. Last images in each row demonstrate that outer bounds of emerged patterns (shown blue) circumscribe corresponding seed patterns (shown red). In the rest of color coded images cells that are to die on the next step are shown pink, cell that are to give a birth are shown green, stable "live" cells are shown blue and "empty" cells are shown black; in black and white images "live" cells are shown white.

One can see from the figures that, whereas growing formations for the Standard_mask model propagate till they reach boundaries of the "vital space" and then stay active in this space, which demonstrates a capability of "unlimited" expansion, dynamics of other models is different. For the Diagonal_mask, Cross_mask, Hex0_mask, Hex1_mask and Hex2_mask models, growing formation reach shapes bounded by certain geometric figures specific for each particular model: rectangle or right angles (Isotropic_mask and Diagonal_mask models), octagon (Cross_mask model), hexagon (Hex0_mask and Hex1_mask models). Remarkably that the process of formation, from chaotic and sufficiently dense seed patterns, of active "communities" stable in size and shape starts from shrinkage of the "vital space" from corners and the shrinkage stops when the "community" reaches a certain "critical" size (Figure 10).

## 6. CONCLUSION

Several modifications of the standard Conway's Game of Life have been suggested and evolutionary dynamics of the introduced new models has been experimentally investigated. In the experiments, a number of new phenomena have been revealed. Specifically

(i) "Ordering of chaos" into maze-like patterns with stochastic "dislocations". These patterns remind patterns of magnetic domains, finger prints, zebra skin, tiger fur, fish skin patterning and alike, which can frequently be found in live as well as in inanimate nature. For the Standard_mask model, these patterns, being stable in the "vital space", preserve a capability "self-healing' and implantation.

(ii) "Self-controlled growth" into "birth/death" - active "communities" bounded by shapes, specific of each model (squares or right angles, octagons, hexagons oriented parallel to the model rectangular lattice axes or $45^{o}$ rotated with respect to the lattice axes)

(iii) "Coherent shrinkage" of the formations to nil or to a few stable or oscillating formations keeping in this process isomorphism of their bounding shapes until the very end.

(iv) "Eternal life in a self-bounded space": permanent "birth-death" activity within "communities" bounded by shapes specific for each model and reached through a process of the "self-controlled growth"